\documentclass{article}
\usepackage{upgreek}
\usepackage{svg}
\usepackage{braket}
\usepackage{lipsum} 
\usepackage{amsmath}
\usepackage{amssymb}
\usepackage[margin=116pt]{geometry}

\begin{document}
\textwidth=380pt
\title{Integrated photonic Galton board and its application for photon counting}

\author{Hezheng Qin$^1$, Risheng Cheng$^1$, Yiyu Zhou$^1$, and Hong X. Tang$^1$*\\
	$^1$Department of Electrical Engineering,\\
	Yale University, New Haven, CT 06520, USA\\	
	*hong.tang@yale.edu}
\maketitle



\begin{abstract}
The Galton board is a desktop probability machine traditionally used to visualize the principles of statistical physics with classical particles. Here, we demonstrate a photonic Galton board that enables on-chip observation of single-photon interference. The photonic Galton board, which can be considered as a simplified Boson sampler, consists of a directional coupler matrix terminated by an array of superconducting nanowire detectors to provide spatiotemporal resolution. This design also allows for photon-number-resolving capability, making it suitable for high-speed photon counting. Our results demonstrate the compatibility between single-photon detector array and photonic integrated circuits, paving the way for implementing on-chip large-scale quantum optics experiments and photonic quantum computing.
\end{abstract}


\section{Introduction}
Quantum computing promises to solve complex problems with unprecedented speed by performing data processing with qubits~\cite{ekert1996quantum}. Among various types of qubits, photons stand out as a potential candidate due to their long coherence time and natural connectivity to modern optical communication systems~\cite{o2007optical,carolan2015universal}. In general, photonic quantum computing is performed by sending quantum states to a universal linear optical network and measuring output states by single-photon detectors~\cite{knill2001scheme,nielsen2004optical,browne2005resource,maring2024versatile}, and Boson sampling is a nominal example~\cite{tillmann2013experimental,huh2015boson}. While quantum advantages have been achieved by a tabletop Boson sampler~\cite{zhong2020quantum}, the large footprint fundamentally limits further scalability. By contrast, photonic integrated circuits (PICs) provide a compact and low-loss platform for implementing a large array of optical components, making them well-suited for realizing on-chip linear optical networks~\cite{o2009photonic, wang2020integrated,elshaari2020hybrid,alexander2024manufacturable,pelucchi2022potential}. In addition, on-chip thermal and electrical phase shifters can be integrated into photonic integrated circuits, enabling the development of a reconfigurable optical network~\cite{maring2024versatile}. However, the low chip-to-detector coupling efficiency remains a bottleneck that suppresses the computational power boost of PIC-based photonic quantum computers. 

Here, we demonstrate the integration of the directional coupler matrix and the superconducting nanowire single-photon detector (SNSPD) array~\cite{esmaeil2021superconducting,gyger2021reconfigurable,chang2021detecting,ferrari2018waveguide} on a silicon nitride (SiN) waveguide. This integrated design eliminates the chip-to-detector coupling loss and thus represents a significant advance in device architecture for large-scale quantum optics experiments. By injecting coherent states into the input port, our device simplifies to a photonic Galton board, which has been commonly used for quantum walk experiments~\cite{aharonov1993quantum, childs2009universal,grafe2016integrated}. Unlike classical Galton boards that generate a binomial probability distribution, photonic Galton boards produce intricate fringes in the distribution at the output ports that do not converge to a continuous function even after long-time averaging, which is a signature of the quantum interference of photons~\cite{peruzzo2010quantum,simon2020quantum}. The photonic Galton boards have been experimentally realized by free-space optics~\cite{broome2010discrete, kitagawa2012observation, cardano2015quantum}. Alternative demonstrations in the time and frequency domain~\cite{schreiber2010photons, wimmer2018linear, weidemann2022topological, guan2023tera, bouwmeester1999optical} have also been reported, where scalability is sacrificed to achieve detection efficiency. By contrast, our design shows no such trade-off by taking advantage of the inherent scalability of PICs. Additionally, we show that our device can also be used as a high-speed photon counter with photon-number-resolving capability. The directional coupler array splits the input photons to 16 paths, and each path is terminated by a SNSPD. The output electrical pulses from each SNSPD are distinguishable in the time domain by inserting microwave delay lines between neighboring SNSPDs. Therefore, our device can resolve the input photon number by counting the number of output electrical pulses.

 \begin{figure}[]
 \centering\includegraphics[width=\textwidth]{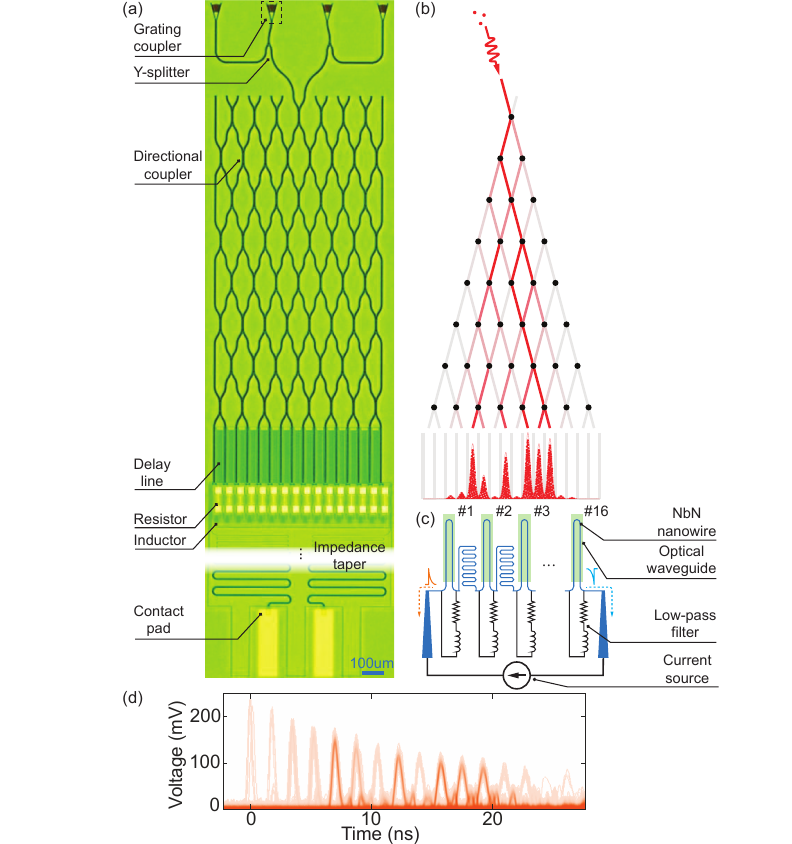}
 \caption{\small The photonic Galton board layout and its working principle.
 (a) Optical micrograph of the device. The dashed frame marks the input grating coupler. For better visibility, the image is taken before the aluminum ground plane deposition (defined by the light green area).
 (b) Schematic illustration of photonic Galton board. Black pegs and gray paths sketch out the directional couplers and optical waveguides. Color intensity on the paths indicates the probability of photons passing them. The photon number in bins shows the distribution formed by single photon input from quantum interference.
 (c) Schematic of SNSPD array for spatiotemporally resolving single photons. Nanowires are biased with a DC current, and microwave delay lines (blue color) are inserted between neighboring nanowires. Each nanowire is shunted by a resistor and an inductor (black color) to form a low-frequency local reset loop. This design can produce both a positive and a negative pulse to determine the position and time of the incident photon.
 (d) Persistence oscilloscope traces of electrical pulses generated from each detector. 16 discrete peaks tell the position of fired nanowires, and the intensity reflects the probability of that pixel being fired.
 }
 \label{fig:galton_board}
 \end{figure}

\section{Design and Simulation}

Our photonic Galton board design resembles that of a classical mechanical Galton board, where billiards percolate through pegs and are collected by bins. As shown in Fig. \ref{fig:galton_board}(a), photons enter the SiN waveguide from a fiber coupled to the input grating coupler. The waveguide is then split into two branches by a Y splitter. The left branch is connected to an output grating coupler for alignment optimization and power calibration, and the right branch guides photons to the photonic Galton board area, comprising 8 stages of directional couplers and then terminated by 16 nanowires. As the analogy shows in Fig. \ref{fig:galton_board}(b), photons can be thought of as billiards, and each directional coupler works as a peg in a classical Galton board, while each single-photon detector pixel counts photons as a bin that collects billiards. All waveguides have a 1.2~$\mu$m width and a 0.4~$\mu$m coupling gap. Nanowires are made of 80-nm-wide, 8-nm-thick, U-shaped niobium nitride (NbN) films extending 80~$\mu$m above waveguides to guarantee full absorption near the 1550~nm wavelength~\cite{pernice2012high,cheng2023100}. Each U-shaped nanowire is shunted by a 5~$\mu$m $\times$ 50~$\mu$m Cr/Au film resistor and an inductor made of 400-nm-wide NbN meander wire to reset bias current after firing. Adjacent pixels are connected by an 800-nm-wide NbN meander delay line, as illustrated in Fig.~\ref{fig:galton_board}(c). The two boundary pixels are connected to the impedance tapers made of meandered NbN wires with a gradually varying width from 800 nm to 9.2 $\mu$m to match the 50~$\Omega$ impedance of the readout electronics. To distinguish photon-triggered electrical pulses from different nanowires, all parts of the delay line are covered by a high-dielectric-constant AlO$_x$ layer and an aluminum (Al) top ground layer to form a low-speed microstrip line, which results in a 0.9~ns delay time and thus helps distinguish microwave pulses generated by neighboring detector pixels in the time domain. 

When a photon is absorbed by a DC-biased superconducting nanowire, it generates a local resistive hotspot that subsequently recovers, producing both positive and negative electrical pulses. The negative pulse travels along the direction of the DC current on the microwave bus line, while the positive pulse propagates in the opposite direction. The arrival time difference of the positive and negative pulse pair is used to distinguish photon position in bins, which varies by 1.8~ns between neighboring bins. The output of such a photonic Galton board system can be visualized by the persistence trace on the oscilloscope in Fig.~\ref{fig:galton_board}(d), where the negative pulses are used as the trigger, and the positive pulses are plotted. These pulses show well-distinguished peaks, while the intensity of the persistence trace corresponds to the number of photons absorbed by each nanowire. We note that the pulse amplitude decreases at a larger time delay due to the propagation attenuation.

\begin{figure}[!t]
 \centering\includegraphics[width=\textwidth]{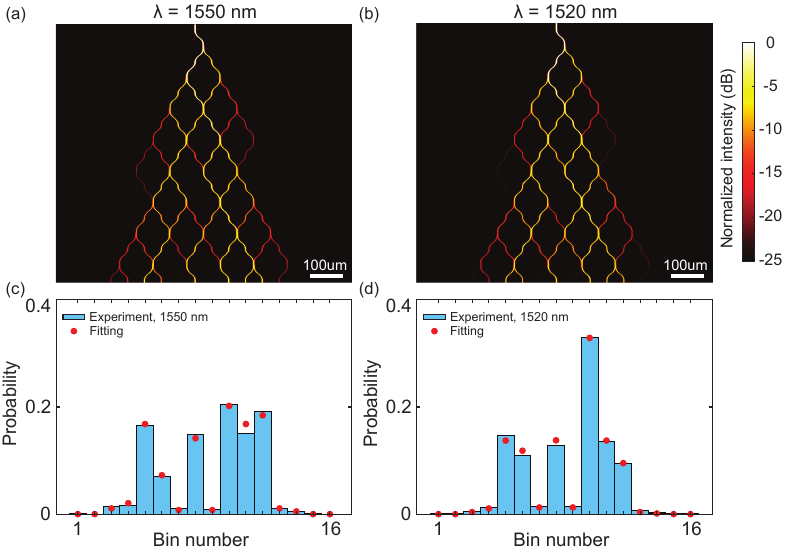}
 \caption{\small Simulation and calculation applied on photonic Galton board. The left panels correspond to 1550~nm photons, and the right panels correspond to 1520~nm photons. 
 (a-b) The FDTD simulation on the directional coupler array shows how a photon interferes with itself and forms the fringes at the output ports. 
 The picture is compressed vertically by a factor of two to reduce the figure size, and the waveguide area is magnified to better visualize intensity distribution. (c-d) Calculated probability distribution by iteratively applying creation operators is used to fit the experiment results with the least square method. The blue bars are the single-photon probability distribution measured in the experiment; they agree well with the red dots following the theoretical distribution. The fitting results are $t^2=0.763\pm 0.003$ for 1550nm photons and $t^2 = 0.816\pm 0.003$ for 1520nm photons. Fitting parameters are reported as least squares estimators and the 95$\%$ confidence intervals.}
 \label{fig: design simulation}
 \end{figure}

The propagation of a single photon in the photonic Galton board follows a discrete-time quantum walk model~\cite{grafe2016integrated}. Assuming $\hat{a}^\dagger_{L}$ and $\hat{a}^\dagger_{R}$ correspond to the creation operator in the left and right branches respectively, each directional coupler can be modeled as
\begin{equation}
\begin{pmatrix}
\hat{a}^\dagger_{L}\\
\hat{a}^\dagger_{R}
\end{pmatrix}_{\textrm{out}}
=
\begin{pmatrix}
t&ir\\
ir&t\\
\end{pmatrix}
\begin{pmatrix}
\hat{a}^\dagger_{L}\\
\hat{a}^\dagger_{R}
\end{pmatrix}_{\textrm{in}},
\label{eq:beam_splitter}
\end{equation}
where $t$ and $r$ are the transmission and coupling ratios in direction couplers and follow the relation of $t^2+r^2=1$. The probability distribution of the photons at the final output ports can be derived by iteratively applying the relation. For our devices, eight iterations are applied to generate a single-photon probability distribution in the 16 bins of photonic Galton board. We treat $t^2$ as a free parameter to fit the single-photon probability distribution to the experimentally determined probability distribution, as illustrated in Fig.~\ref{fig: design simulation}(c-d). The fitting results are $t^2=0.763\pm 0.003$ for 1550~nm photons and $t^2 = 0.816\pm 0.003$ for 1520~nm photons. Fitting parameters are reported as least squares estimators and the 95$\%$ confidence intervals. It is noteworthy that such a result can also be obtained by classical wave optics if we interpret the light wave intensity as the probability of a photon, which is evidence for the wave-particle duality of photons~\cite{grafe2016integrated}. Therefore, the effect of the directional coupler array can be simulated by the finite-difference time-domain (FDTD) method, showing how a photon interferes with itself and generates a spatial probability distribution at the output port. We apply the directional coupler geometry parameters to the FDTD model and obtain $t^2=0.768$ for 1550 nm light and $t^2 = 0.804$ for 1520 nm light, and the corresponding simulation results are shown in Fig.~\ref{fig: design simulation}(a-b).

\section{Device fabrication and results}
The fabrication starts with a 330-nm-thick SiN film on 3.3-$\mu$m-thick thermal oxide on a silicon wafer, and the SiN is deposited by low-pressure chemical vapor deposition. An 8-nm-thick NbN superconducting thin film is deposited on the SiN film by the atomic layer deposition. All pattern exposures are performed by electron beam lithography. The NbN film is etched by the tetrafluoromethane (CF$_4$) gas using the 6$\%$ hydrogen silsesquioxane (HSQ) as the resist, and the SiN film is etched by the fluoroform (CHF$_3$) using AR-P 6200 (CSAR 62) as the resist. The commonly used double-layer polymethyl methacrylate (PMMA) lift-off process and electron beam evaporation are employed to pattern all other layers, such as Cr resistors, AlO$_x$ dielectric spacers, and aluminum top grounds.

\begin{figure}[ht]
 \centering\includegraphics[width=\textwidth]{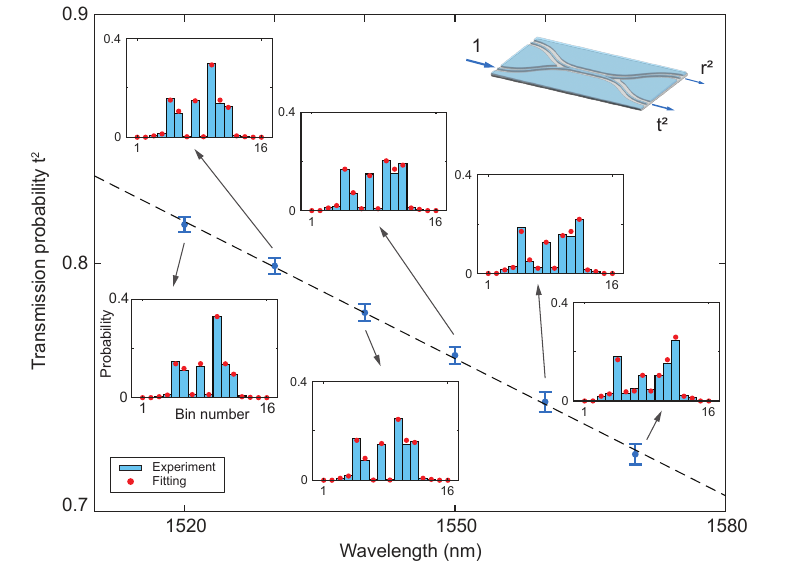}
 \caption{\small At different wavelengths, quantum interference is observed at output ports when a single photon enters the input port. This result suggests the wavelength-dependent transmission probability $t^2$ through a directional coupler. The wavelength dependence is fitted by a linear dashed line, indicating $t^2$ through a directional coupler change linearly with wavelength within our measurement range. 
 (Insets) The experimental data at different wavelengths represented by the blue bars matches well with the theoretical distribution, indicated by the red dots fitted using the least square method. All axis labels are the same as the first one. The sample size used in all insets is 10,000. Data points and error bars reflect least squares estimators and 95$\%$ confidence intervals.}
 \label{fig:single_stat}
 \end{figure}
 
We first demonstrate the performance as a photonic Galton board by measuring quantum interference at output ports when a weak coherent state enters the input port. An oscilloscope is used to collect electrical pulses that are generated from detectors. To determine the arriving time interval between the positive and negative pulses, the negative pulses are used as a trigger for the oscilloscope, and the positive pulse traces are recorded as mentioned above. An attenuated coherent laser source near 1550~nm is used as a weak coherent state source. The laser source is attenuated to a level that the expectation value of the photon arrival time interval $\tau$ is much larger than the recording time window of the oscilloscope. Hence, there is a low probability that more than two photons enter the Galton board and are collected in the same oscilloscope trace. The statistic of the output position of photons can be extracted and fitted as the insets in Fig.~\ref{fig:single_stat} by adjusting the parameter $t^2$. When the wavelength of a photon is tuned, the statistical probability changes accordingly (Fig.~\ref{fig:single_stat}). This result reveals a wavelength-dependent transmission $t^2$ through a directional coupler. It can be seen that $t^2$ increases linearly with wavelength within our measurement range, which agrees with the FDTD simulation.

 \begin{figure}[ht]
 \centering\includegraphics[width=\textwidth]{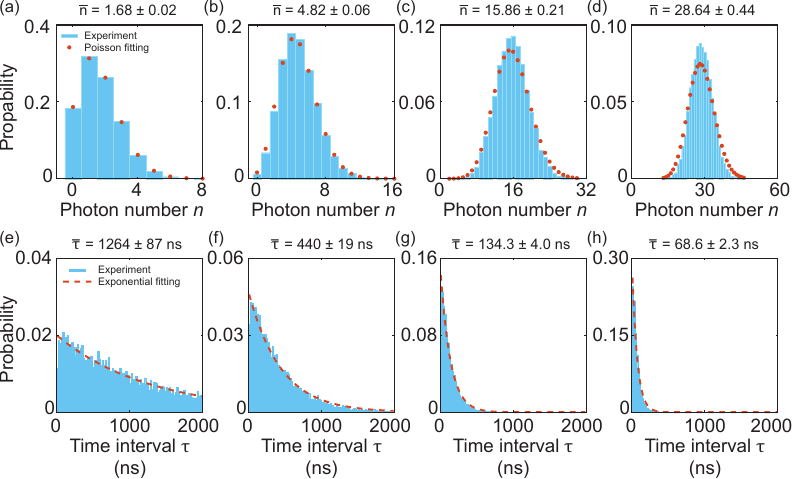}
\caption{\small Photonic Galton board system reveals photon number statistics. All upper panels show the results from directly counting photon numbers in a time window, and all lower panels show the results from measuring time intervals between adjacent photons. Panels in the same column have the same light intensity to compare the two methods. The blue bars are the experimental data when coherent light is used as the light source; these agree well with the red dots or dashed lines following the theoretical distribution. The sample size in all figures is 10,000. Fitting parameters are presented as least square estimators and $95\%$ confidence intervals. The parameters $\bar{n}$ and $\bar{\tau}$ represent the mean values of the photon number $n$ and time interval $\tau$, respectively.} 
 \label{fig:multi_stat}
 \end{figure}
 
Next, we show that the photonic Galton board integrated with SNSPD can also be used as a 16-pixel spatiotemporally multiplexed photon-number-resolving detector. We first measure the number of output pulses in a time window of $\Delta T=2\mu\textrm{s}$ for direct photon number counting. The directional couplers split photons into different bins to avoid multiple photons firing the same nanowire within its reset time. After collecting a sufficient number of measurements, a histogram is constructed to show photon number statistics (Fig.~\ref{fig:multi_stat}(a-d)). It can be seen that the experimental data agree well with the Poisson fitting $P(n;\bar{n})=\bar{n}^n\textrm{exp}(-\bar{n})/n\textrm{!}$, where the fitted mean photon number $\bar{n}$ is represented by least square estimators and $95\%$ confidence intervals. We note that the detector is able to count more than 16 photons because the detection time window $\Delta T$ is much larger than the nanosecond-level nanowire reset time. However, as the average photon number grows larger than the detecting pixel number, the saturation effect (i.e., a single pixel is fired by multiple photons within its reset time~\cite{cheng2023100}) is still observable, as the measured histogram starts to deviate from a Poisson distribution (see Fig.~\ref{fig:multi_stat}(d)).

Alternatively, for a Poisson distribution, there is a one-to-one correspondence between the photon number distribution and the time interval distribution~\cite{davenport1958introduction}, and we next show that our Galton board design also allows for direct measurement of the time interval distribution thanks to its temporal resolution. To measure the time intervals between adjacent photons, both negative and positive electrical pulses need to be recorded, and the average arrival time of the positive and negative pulse is used to determine the arrival time of a photon. The traveling time difference between positive and negative electric pulses on microwave delay lines is thus averaged out~\cite{oripov2023superconducting}. The histogram of the time interval is shown in Fig.~\ref{fig:multi_stat}(e-h). The statistical data is fitted by a negative exponential distribution $P(\tau;\bar{\tau}) \propto\textrm{exp}(-\tau/\bar{\tau})$ for coherent light. It can be seen that the experimental data agree well with the exponential fitting. In this method, the average photon number revealed by time interval statistics is $\bar{n} = \Delta T/\bar{\tau}$, where $\bar{\tau}$ is the decay time constant of the exponential distribution~\cite{davenport1958introduction}. This time interval measurement can be useful for applications such as time-resolved fluorescence measurements~\cite{badea197917,soini1987time} and correlation measurements~\cite{glauber1963photon}.

\section{Conclusion}
In this work, we report a photonic circuit integrated with an on-chip SNSPD array to realize the photonic Galton board. Our device is able to measure the photon number received by each nanowire, which reveals the single-photon interference effect occurring in the directional coupler matrix. Besides, our design also allows us to measure both photon number probability distribution and time interval distribution, thanks to its photon number resolution and time-resolving capability. We note that the photon number resolution comes from the spatially multiplexed structure of the photonic Galton board, while the time-resolved measurement benefits from the fast reset time of nanowires. We believe it is feasible to further scale up the directional coupler matrix and superconducting pixels and operate on a larger photon number basis. More complex optical components, such as phase shifters and entangled photon sources, can also be integrated into the platform to facilitate the implementation of reconfigurable large-scale quantum optics experiments and photonic quantum computing.

\section*{Funding}
This work is funded in part by the Department of Energy under grant No. DE-SC0019406. The materials used in this work is developed under the support of the Office of Naval Research (ONR) grant N00014-20-1-2126.

\section*{Acknowledgments}
The authors thanks Michael Rooks, Yong Sun, Lauren McCabe and Kelly Woods for support in the cleanroom and assistance in device fabrication. 

\section*{Disclosures}
The authors declare no conflicts of interest.

\section*{Data availability} Data underlying the results presented in this paper are not publicly available but may be obtained from the corresponding author upon reasonable request.

\bibliographystyle{unsrt}
\bibliography{references}
\end{document}